\documentclass[article,prl,twocolumn,nofootinbib,superscriptaddress]{revtex4-1}

\usepackage{amsmath,xcolor,amsfonts,amsthm,amssymb,graphicx, srcltx,hyperref}
\usepackage{times}

\newcommand{\ket} [1] {|#1\rangle}

\begin{document}

\title{Experimental Satellite Quantum Communications}
\author{Giuseppe Vallone}
\author{Davide Bacco}
\author{Daniele Dequal}
\author{Simone Gaiarin}
\affiliation{Dipartimento di Ingegneria dell'Informazione, Universit\`a degli Studi di Padova, Padova, Italy}
\author{Vincenza Luceri}
\affiliation{e-GEOS spa, Matera, Italy}
\author{Giuseppe Bianco}
\affiliation{Matera Laser Ranging Observatory, Agenzia Spaziale Italiana, Matera, Italy}
\author{Paolo Villoresi}
\email{paolo.villoresi@dei.unipd.it.}
\affiliation{Dipartimento di Ingegneria dell'Informazione, Universit\`a degli Studi di Padova, Padova, Italy}

\begin{abstract}
Quantum Communications on planetary scale require complementary channels including ground and satellite links. The former have progressed up to commercial stage using fiber-cables, while for satellite links, the absence of terminals in orbit has impaired theirs development.
However, the demonstration of the feasibility of such links is crucial for designing space payloads and to eventually enable the realization of protocols such as quantum-key-distribution (QKD) and quantum teleportation along satellite-to-ground or intersatellite links.
We demonstrated the faithful transmission of qubits from space to ground by exploiting satellite corner cube retroreflectors acting as transmitter in orbit, obtaining a low error rate suitable for QKD.
We also propose a two-way QKD protocol exploiting modulated retroreflectors that necessitates a minimal payload on satellite, thus facilitating the expansion of Space Quantum Communications.
\end{abstract}

\maketitle

Quantum Physics protocols connecting separate operators call for Quantum Communications (QCs), whose essence is the fair transport of a generic quantum state along a channel \cite{vane98sci}.
For their implementation, quantum protocols may require short links, as in the case of Quantum Computing, or extended connections, as for the realization of Quantum Key Distribution (QKD), in which two distant parties generate a secret key to be used for data encryption \cite{benn99sci, lo99sci, hugh11sci, bacc13nco}. In the case of QKD, the locations of these parties may be connected by optical cables, for which its implementation has already reached its maturity \cite{peev09njp,sasa11ope,froh13nat}, or by two ground optical terminals \cite{ursi07nap}. However, a significant and crucial fraction of QKD applications require a free-space network including one or more nodes on a satellite, as it is occurring for classical communications (e.g. internet, phone and TV). More generally, Space QCs are needed to extend fundamental tests of Quantum Mechanics as the measure of Bell's inequality in a relativistic scenario, as well as to establish a global network to distribute entanglement and provide secure communications \cite{legg09, ride12cqg}, as fostered in several continental Information-and-Communication-Technology (ICT) roadmaps \cite{xin11sci, roadmaps}.
\begin{figure}[t!]
\begin{center}
\includegraphics[width=5.5cm]{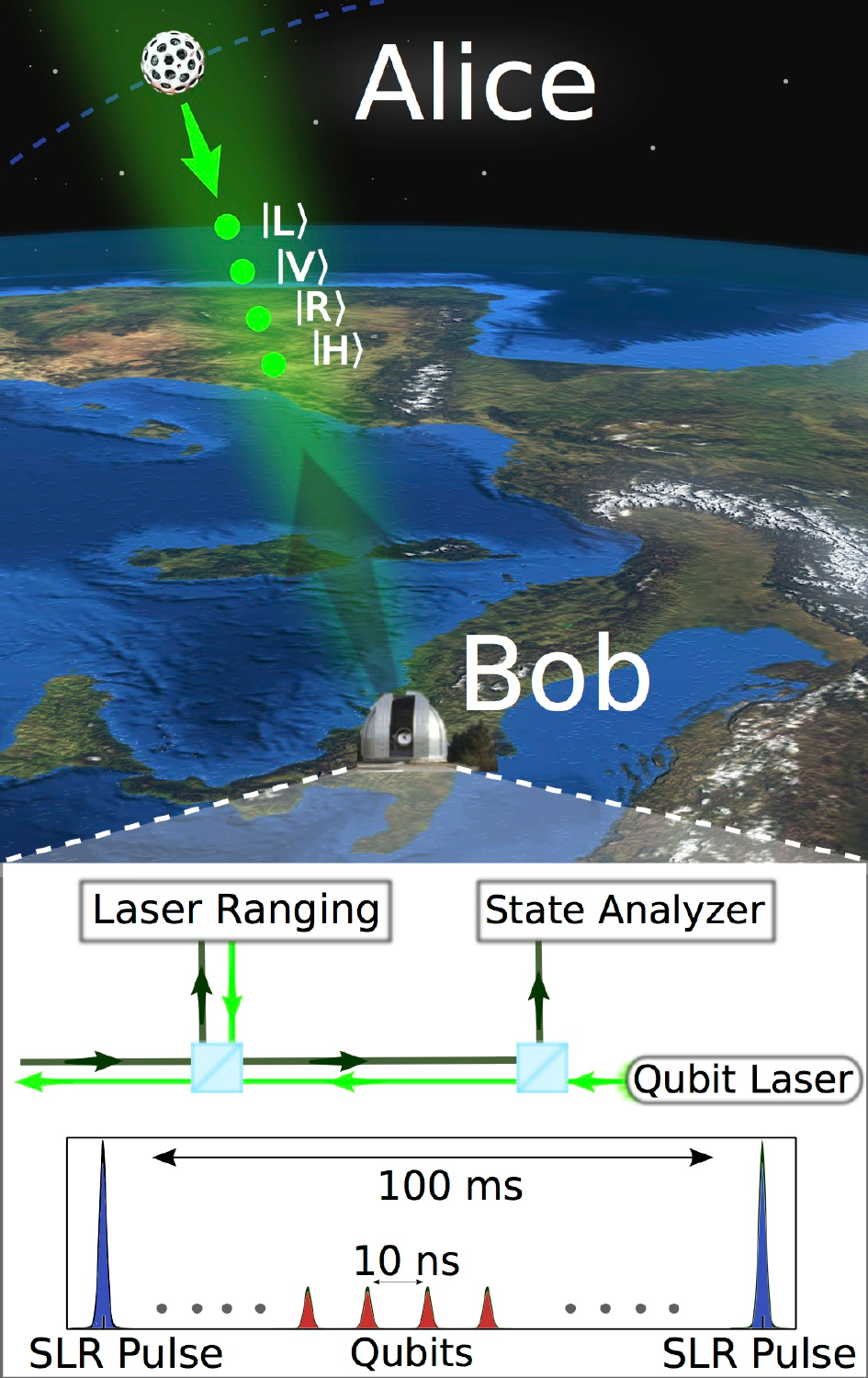}
\caption{Scheme of the Satellite QKD demonstration. Qubit pulses are sent at $100$ Mhz repetition rate and are reflected back at the single photon level from the satellite, thus mimicking a QKD source on Space. Synchronization is performed by using the bright SLR pulses at repetition rate of $10$ Hz.}
\label{fig:setup}
\end{center}
\end{figure}

The envisaging and modelling of Space QCs started a dozen years ago \cite{aspe03ieee, bona06ope, toma11asr, bour13njp,sche13njp}. Nevertheless, sources or detectors of quantum states have not been placed in orbit yet. Therefore, since 2008 the experimental studies of Space-to-ground links simulated a source of coherent pulses attenuated at the single photon level by exploiting satellites for geodetics laser ranging, which are equipped with corner-cube retroreflectors (CCRs) \cite{vill08njp, yin13ope}.  However, a full quantum transmitter for polarization encoded QKD in Space also requires qubits prepared in different polarization states.
\begin{figure*}[t!]
\begin{center}
\includegraphics[width=14cm]{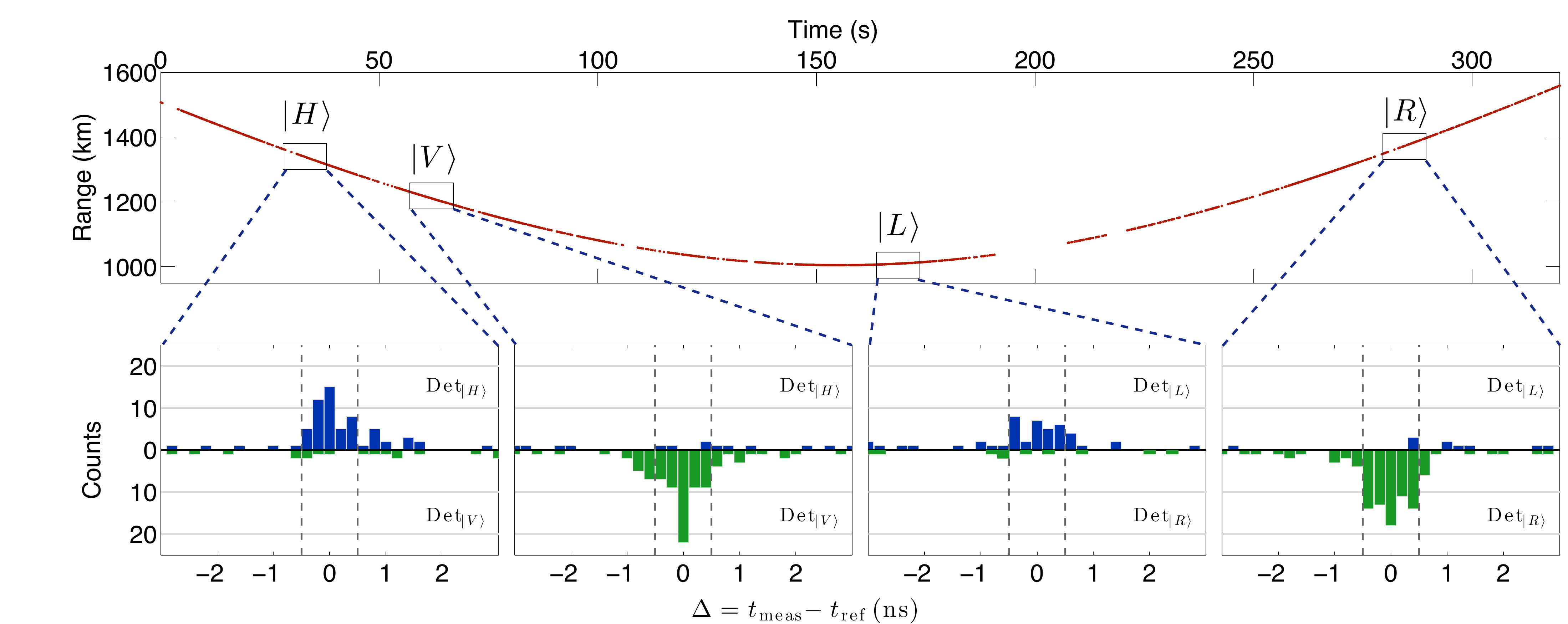}
\caption{Top: Larets trajectory measured by the $10$ Hz SLR pulses. The four selected $10$ s intervals correspond to four different polarization input states. Bottom: the four histograms report the obtained counts at the receiver for each single photon detector in function of the measured detection time $t_{\rm meas}$, demonstrating an average QBER of $6.5$ \%. The signal on the two detectors is blue for H/L polarization and green for V/R. Gray dashed lines represent the $1 \, \sigma$ selection interval around the expected time of arrival $t_{\rm ref}$.}
\label{fig:larets}
\end{center}
\end{figure*}

Here we show the operation of such quantum transmitter by sending toward selected satellites a train of laser pulses at the repetition rate of $100$ MHz paced with an atomic clock and generated at the Matera Laser Ranging Observatory (MLRO) of the Italian Space Agency in Matera (Italy). The qubit signal is obtained by the pulses reflected by the CCRs (see \figurename~\ref{fig:setup}).
We set the outgoing laser intensity such that the qubit signal has an average photon number per pulse $\mu_{sat}$ close to one, as required by QKD protocols. Then we demonstrate the faithful transmission of different polarization qubits between satellite and ground and provide a quantitative estimate of the mean photon number of the downward pulses. We will show at the end that, by using modulated CCRs, our technique can be exploited for a full QKD protocol along Space channels. Our results prove that quantum key distribution from an orbiting terminal and a base station is not only a promising idea but is nowadays realizable.

We prove the feasibility of the BB84 protocol \cite{BB84} with the qubits encoded in four different polarization states, corresponding to two mutually unbiased basis. A secret key can be established between the transmitter (Alice, at the satellite) and the receiver (Bob, at MLRO) when the average Quantum Bit Error Rate (QBER) is below $11$\% \footnote{By using the post-selection techniques introduced in \cite{bae07pra}, QBER up to $20$\% can be tolerated for secret key generation.}. In a transmission with polarization qubits, the QBER can be estimated as
$Q=(n_{\rm wrong}+1)/(n_{\rm corr}+n_{\rm wrong}+2)$
where $n_{\rm corr}$ and $n_{\rm wrong}$ are the number of detections in the sent and orthogonal polarization respectively \footnote{We used the Bayesian estimator of the QBER.}.
The exploitation of CCRs with metallic coating on the three reflecting faces is crucial for preserving the imposed polarization state during the reflection. For this reason we could not use satellites mounting uncoated or dielectric coated CCRs. We selected five LEO (Low Earth Orbit, below $2\,000$ km) satellites: Jason-2, Larets, Starlette and Stella with metallic coated CCRs and Ajisai, with uncoated CCRs, for comparison.

In order to reject the background and dark counts, a precise synchronization at Bob is needed. For this purpose we exploited the satellite laser ranging (SLR) signal. The latter is generated in a much coarser comb of strong pulses ($10$ Hz repetition rate and $100$ mJ pulse energy) whose seed is taken from the same comb used for the qubits. Two non-polarizing beam splitters were used in the optical path in order to merge and split the outgoing and incoming SLR signal and qubit stream (see \figurename~\ref{fig:setup}). For qubits discrimination, we synchronized the state analyser with the time-tagging of SLR pulses provided by the MLRO unit, which has few picosecond accuracy. Indeed, by dividing the intervals between two consecutive SLR detections in $10^7$ equidistant subintervals, we determined the sequence of expected qubit times of arrival $t_{\rm ref}$. This technique compensates for the time scale transformation due to satellite motion with respect to the ground. Our detection accuracy $\sigma$ was set equal to the detector time jitter ($0.5$ ns), as other contributions to time uncertainties coming from detection electronics or laser fluctuations are negligible. Counts registered within $1 \, \sigma$ interval around $t_{\rm ref}$ were considered as signal, while the background is estimated from the counts outside $3 \, \sigma$. Details of the setup are described in Supplementary Material.

\paragraph*{QCs of polarized photons} QCs using generic polarization states from two mutually unbiased bases were realised with a single passage of Larets. The passage was divided in four intervals of $10$ s in which we sent horizontal $\ket{H}$, vertical $\ket{V}$, circular left $\ket{L}$ and circular right $\ket{R}$ states. At the receiver the state analysis is performed by two single photon detectors measuring two orthogonal polarizations, from which the QBER is extracted. The results are summarized in \figurename~\ref{fig:larets}. In the four intervals, we obtained $199$ counts in the correct detector and $13$ wrong counts, giving an average QBER of $6.5$ \%. Once considered the average $3.6$ \% duty cycle of our setup (see Supplementary Material), the mean return frequency in the selected intervals is $147$ Hz.

\begin{figure}[t!]
\begin{center}
\includegraphics[width=7cm]{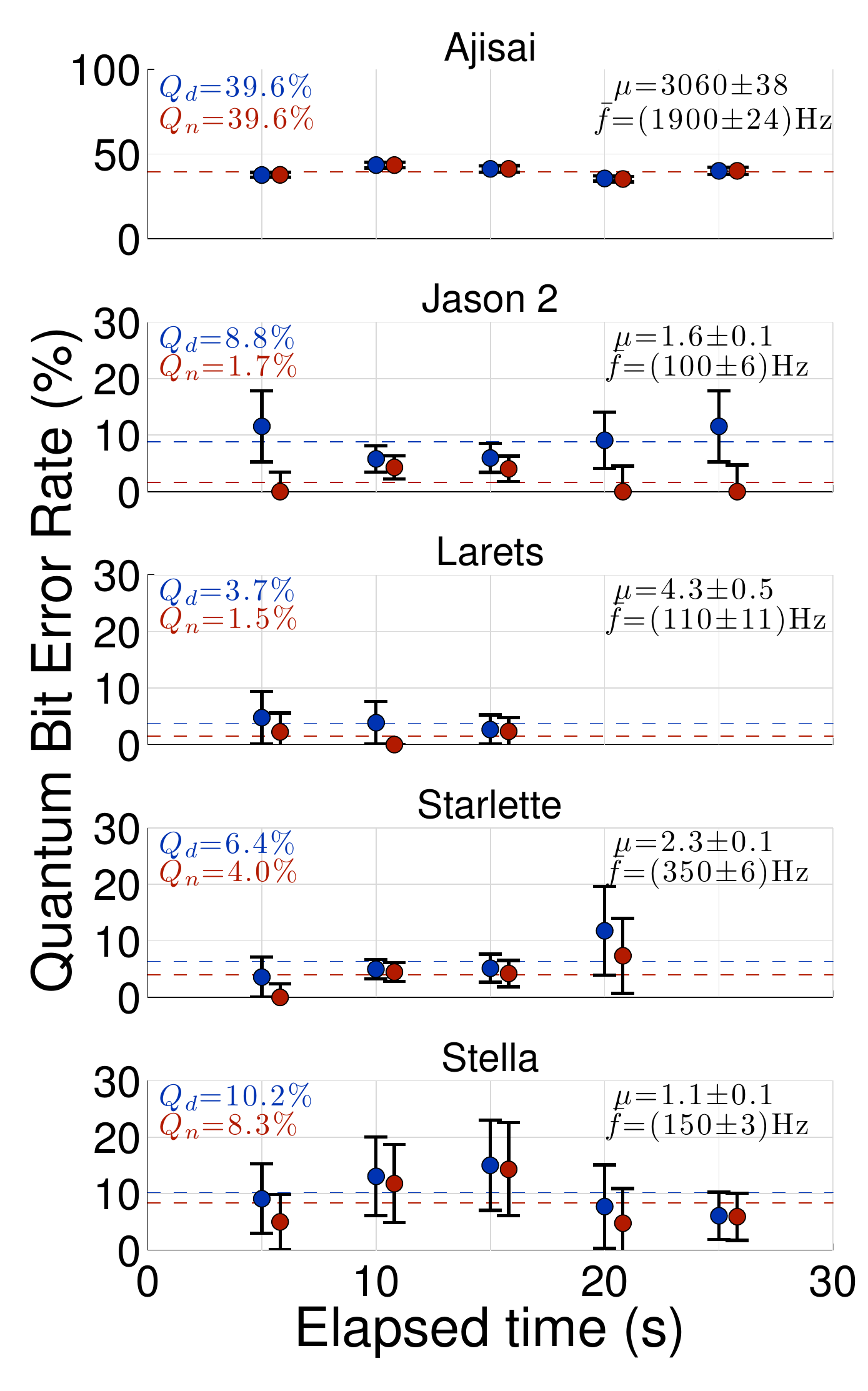}
\caption{Quantum Bit Error Rate (QBER) of the signal received from different satellites. We fixed the sent polarization to $\ket{V}$ and measured in two orthogonal polarization $\ket{H}$ and $\ket{V}$. For each satellite we show the bare QBER ($Q_d$) and the QBER obtained by subtracting the background ($Q_n$). Error bars represent Poissonian errors. The coating of Ajisai retroreflectors depolarizes the qubits, while the other satellites preserve the photon polarization. We also indicate the mean detection rate and the average photon number per pulse at the satellite.}
\label{fig:QBER}
\end{center}
\end{figure}

A further analysis has been carried out to prove the preservation of the polarization state for the other coated satellites. These results will prove that low QBERs can be obtained in different conditions and satellite orbit, showing the stability and the reliability of our approach. We will also report the detection rates achievable with the different LEO satellites. In this analysis we divided the detection period in intervals of $5$ seconds: for each interval the data were analyzed only if the signal of at least one detector was $5$ standard deviations above the background. The QBERs resulting from this analysis are shown in \figurename~\ref{fig:QBER} for Ajisai, having non polarization preserving CCRs, and for the polarization preserving satellites Jason-2, Larets, Starlette and Stella. We achieved a QBER below $10$ \% for several tens of seconds in all the polarization maintaining satellites.

\begin{figure}[t!]
\begin{center}
\includegraphics[width=7cm]{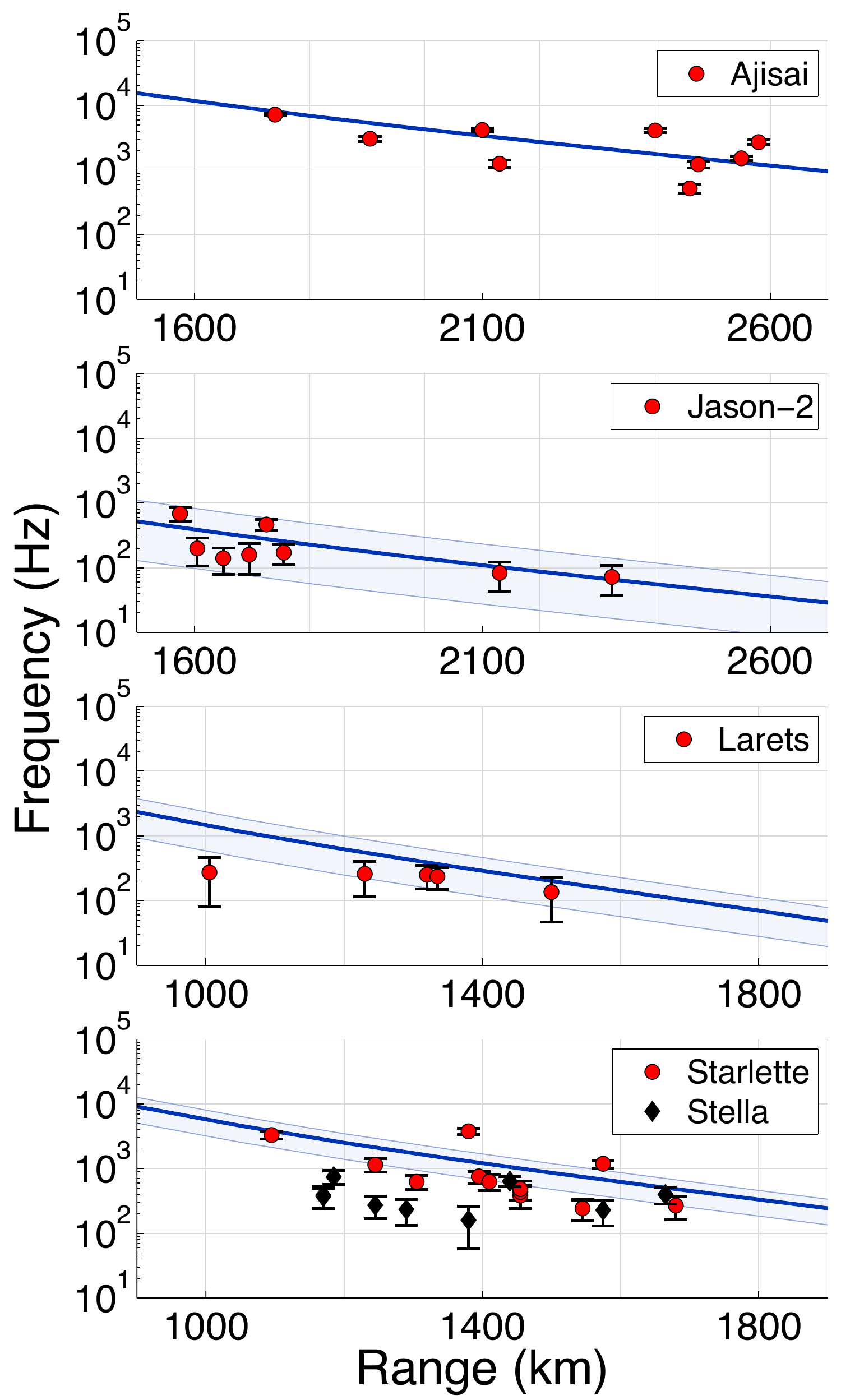}
 \caption{Return frequencies and link budgets. Points represent the return frequencies of the qubits for different satellites along the orbit, compared with the prediction of the link budget provided by the radar equation model (continuous line). Error bars account for Poissonian errors only while shaded area comes from the available uncertainties of the satellite cross-sections $\Sigma$. Uncertainties in the orbital parameters and beam pointing affect trend of the return rate beyond shot noise. The twin satellites Stella and Starlette show different behaviour despite similar characteristics, but in line with the SLR statistics.}
\label{fig:LinkB}
\end{center}
\end{figure}

\paragraph*{Link budget and $\mu$ estimate}
A real earth-satellite QKD system is based on faint laser pulses with a mean photon number of the Poisson process $\mu$ close to $1$. Indeed the BB84 protocol with decoy states \cite{ma05pra,lim14pra} in a realistic scenario requires $\mu\lesssim 2$ \cite{scar09rmp}. We demonstrate in the following that our experiment was carried out in this regime.

An estimation of the average number of photons per pulse leaving the satellites $\mu_{sat}$ is obtained by dividing the average number of photons per pulse detected at the receiver, $\mu_{rx}$,
by the transmission of the quantum channel. A general formula to predict the detected number of photons per pulse is the radar equation
$\mu_{rx} = \mu_{tx} \, \eta_{tx}\, G_t \,\Sigma \left(\dfrac{1}{4 \pi R^2}\right)^2\,T^2_a\,
A_t\, \eta_{rx}\, \,\eta_{det}$
where $\mu_{tx} $ is the source mean photon per pulse, $\eta_{tx}$
is the optical transmission efficiency, $G_t $ is the transmission gain, $\Sigma $ and  $R$ the satellite cross-section and slant distance, $T_a$ the atmospheric transmissivity, $A_t$ the telescope area, $\eta_{rx}$ the optical receiving efficiency and $\eta_{det}$ the single photon detector efficiency \cite{degn93geo}. To estimate $\mu_{sat}$ it is necessary to factorise the radar equation into an uplink and a downlink factors. While most of the parameters of the radar equation can be easily separated into uplink and downlink factors, the satellite cross section $\Sigma$ plays a role in both and must be split according to
$\Sigma = \rho A_{\rm eff} G_{down}$. The parameters $\rho$ and $A_{\rm eff}$, corresponding to the CCR reflectivity and the effective satellite retroreflective area, contribute to the uplink, while $G_{down}$ gathers all the downlink contributions into an effective downlink gain.
Then, the downlink factor, corresponding to the quantum channel transmission, is given by
\begin{equation}
\mu_{rx} = \mu_{sat} \frac{\Sigma}{\rho A_{\rm eff}} \left ( \dfrac{1}{4 \pi R^2} \right ) T_a A_t \eta_{rx} \eta_{det}
\label{eq:dlradar}
\end{equation}
The values of $\mu_{sat}$ in \figurename~\ref{fig:QBER} were calculated by using \eqref{eq:dlradar} with the following parameters $\eta_{det} = 0.1$, $P_s = 0.11 W$, $\eta_{tx} = 0.1$, $A_t = 1.73 m^2$, $\eta_{rx} = 0.13$. We used the satellite cross-section reported in \cite{arno03rep,vasilev,burm04wls}. For the Larets passage of \figurename~\ref{fig:larets} we obtained $\mu_{sat}=3.4\pm0.2$ and a corresponding downlink transmissivity of $\sim4.3\cdot 10^{-7}$ ($63$ dB of attenuation). The resulting $\mu_{sat}$ is of the order of unity for the four satellites with metallic CCRs. The reflectivity $\rho$ of these latter was taken as unitary, setting a higher bound on $\mu_{sat}$. The atmospheric absorbance is proportional to the air-mass (AM), defined as the optical path length through atmosphere normalized to the zenith. In our model we considered $87$\% of transmissivity at the zenith for all the days. This value refers to good sky conditions \cite{degn93geo} which were effectively selected for the experiment.

For a consistency check of our $\mu_{sat}$ estimation, the full radar equation has been used to extrapolate the transmitter gain $G_t$, which depends on the upward beam divergence, the beam wandering due to the atmospheric turbulence and the pointing errors. By averaging the data obtained in different passages of Ajisai, Jason and Starlette, an effective gain $G_t = 1.1 \times 10^9$
was obtained (see Supplementary Material).  The resulting value of $G_t$ has been used in the radar equation to estimate the number of received photons. The theoretical predictions and the experimental data are compared in \figurename~\ref{fig:LinkB}. The results show that radar equation model \cite{degn93geo} and eq. \eqref{eq:dlradar} provides a precise fit for the measured counts and the $\mu_{sat}$ values derived in \figurename~\ref{fig:QBER}.

\paragraph*{QKD satellite protocol using retroreflectors}
We note that, if the outgoing and incoming beam travel through the same optical path, the polarization transformation induced in the uplink by the telescope movements is compensated in the downlink (see Supplementary Material). Therefore, by transforming the state during the retroreflection, we drive the qubit state received on the ground. On this base, we propose a two-way QKD protocol, working as follows. In the ground station, a horizontal polarized beam is injected in the Coud\'e path and will exit the telescope rotated by an angle depending on the telescope pointing. The beam is directed toward a satellite with CCRs having a Faraday Rotator (FR) mounted at the entrance face. By using the FR it is possible to rotate the returning polarization by a suitable angle $\theta$. In the CCR a suitable attenuator lowers the mean photon number to the single photon level.
A measure of the intensity of the incoming beam is desirable in order to avoid Trojan horse attack \cite{gisi06pra} and to guarantee the security of the protocol.
The retroreflected beam then propagates toward the ground telescope, and thanks to the properties of the Coud\'e path, a polarization rotated by $\theta$ with respect to the horizontal will be received.
By this scheme, a decoy state BB84 protocol can be realised between satellite and ground. The experimental results shown above demonstrate that such protocol is currently realizable using few centimetres CCRs and that the MLRO station is suitable for Space QCs.

In conclusion, QCs were demonstrated experimentally from several satellites acting as quantum transmitter and with MLRO as the receiver. QBER was found low enough to demonstrate the feasibility of quantum information protocols such as QKD along a Space channel. Moreover, we propose that a very simple trusted device in orbit, formed by an active CCR mounted on spacecraft and operated in the two-way scheme may provide a simple alternative to a full space terminal, fostering a faster expansion of QCs around the planet and beyond.

\begin{acknowledgements}
We would like to thank Francesco Schiavone, Giuseppe Nicoletti and the MRLO technical operators and Prof. Cesare Barbieri, Andrea Tomaello and Alberto Dall'Arche (University of  Padova) for the collaboration and support. This work has been carried out within the Strategic-Research-Project QUANTUMFUTURE of the University of Padova and the Strategic-Research-Project QUINTET of the Department of Information Engineering, University of Padova.
\end{acknowledgements}

\onecolumngrid
\appendix

\section{\Large Supplementary Information}

\section*{Setup}

\begin{figure}[h]
\begin{center}
\includegraphics[width=12cm]{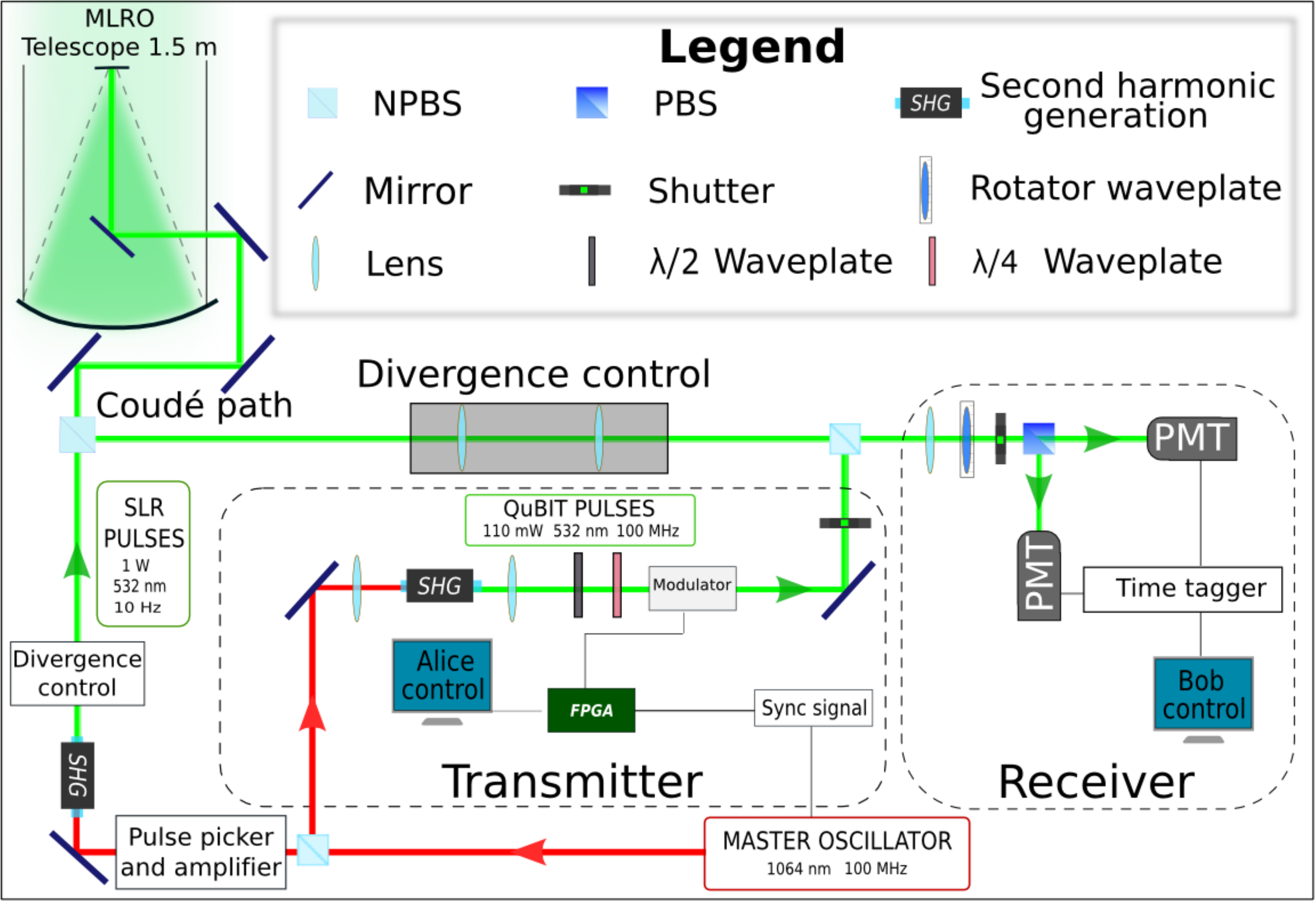}
\caption{Detailed scheme of the experiment.}
\label{fig:setupSI}
\end{center}
\end{figure}

The detailed scheme used in the experiment is shown in figure \ref{fig:setupSI}.

A mode-locking laser oscillator subjugated at the MLRO atomic clock is used as event generator. It produces pulses of $100$ ps duration at the wavelength of $1064$ nm, with $100$ MHz of repetition rate and about $400$ mW of average power. The laser output is split to provide the seed for the SLR signal and the pump pulse for the qubits. The SLR pulse is obtained by selecting with a pulse picker one pulse every $10^7$ and then using a regenerative amplifier and two single-pass amplifiers followed by a Second Harmonic Generation (SHG) stage, obtaining a pulse at $532$ nm with $100$ mJ energy and $10$ Hz repetition rate.
The beam used to generate the qubits is obtained by sending the rest of the master oscillator laser to a suitable SHG unit, whose output is $110$ mW. The beam divergence is controlled by a collimator and the polarization state is controlled by two waveplates and a modulator. Two non-polarizing beam splitters (NPBS) are used to combine SLR with qubit combs in the upward beam that is directed via the Coud\'e path to the MLRO telescope, from which it propagates toward the satellites.

The beams received from the satellites by the MLRO telescope propagate backward via the Coud\'e path and are split by the same two NPBS also used in the uplink. The qubit receiver is composed by a focalizing lens, a rotating waveplate, an optical shutter and two single photon photomultipliers (PMTs) placed at the outputs of a Polarizing Beam Splitter (PBS). The signals detected by the PMTs are fed into a time tagger with $81$ ps resolution. A rotating waveplate, controlled by software, is used to change between two receiving bases, $\{\ket{H},\ket{V}$ and $\{\ket{L},\ket{R}\}$.

\begin{figure}[h]
\begin{center}
\includegraphics[width=7cm]{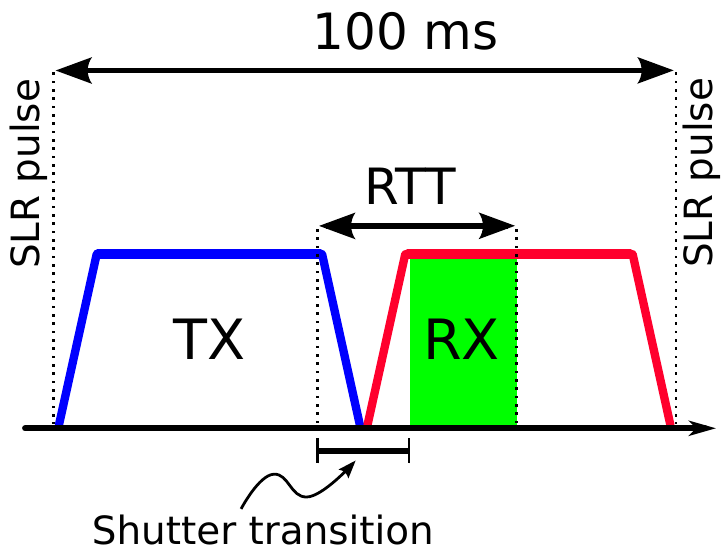}
\caption{Temporal slot between two SLR pulses. TX and RX indicate the transmitter and receiving phase respectively. RTT is the satellite round trip time. The green zone represents the effective time of qubit detection.}
\label{fig:si_time_schedule}
\end{center}
\end{figure}

The pulses generated by the transmitter, passing through the first NPBS produce a scattering that elevates the background noise at the quantum receiver. To prevent this effect, we implemented a time division protocol by using two fast mechanical optical shutters. In the first half of the $100$ ms slot between two SLR pulses, the transmitter shutter is opened in order to send the qubit pulses toward the satellite, while the reception shutter is closed to protect the receiver PMTs. In the second half of the slot the transmitter shutter is closed, and, once the receiver shutter is fully open, the detection phase begin (see \figurename~\ref{fig:si_time_schedule}). By using this protocol, the effective transmission time during a slot cannot be larger than the round trip time (RTT); however, since the shutters require about $2$ ms to fully open and $2.5$ ms to fully close, the effective period is further reduced by $4.5$ ms. Considering that for a LEO satellite the RTT varies between $5$ and $20$ ms, the effective duty cycle can varies between $0$ and $15$ \%. Moreover, the effective duty cycle varies also in a single satellite passage, approaching its minimum when the satellite reaches its maximum elevation.

\section*{Radar equation}
As said in the main text, in order to have an additional confirmation that the obtained values of $\mu_{sat}$ are correct, the full radar equation has been used to extrapolate the transmitter gain, given by
$$ G_t = \dfrac{8}{\theta_t^2} \exp \left [ -2
\left ( \dfrac{\theta}{\theta_t} \right )^2 \right ].  $$
In the previous equation $\theta_t$ is the divergence angle of the upgoing beam (including the beam broadening due to turbulence), while $\theta$ is the pointing error.
Since the two parameters $\theta$ and $\theta_t$ cannot be directly and separately measured, we obtained an estimate for $G_t$ by comparison the data obtained in different passages of the several LEO satellites. As a consequence of the pointing error, the detection frequency of the $100$ MHz laser varies strongly with time, thus producing localized peaks of detection for few tens of seconds, followed by the absence of signal. Because of this several periods of at least $10$ seconds, in which the detection frequency was significantly above the background, have been isolated and
only the peak frequency within these periods has been taken into account. To best approximate of $G_t$, we averaged the most stable data taken for Ajisai, Jason and Starlette, thus
obtaining an effective gain of $G_t = 1.1 \times 10^9$.
This value for $G_t$ has been used in the link budget equation to estimate the received photons frequency and then the fit has been compared with the collected data as shown in \figurename~\ref{fig:LinkB}.

\section*{Polarization compensation in the downlink}
\begin{figure}[h]
\begin{center}
\setlength{\unitlength}{8.5cm}%
\begin{picture}(1,1)%
	\put(0,0){\includegraphics[width=\unitlength]{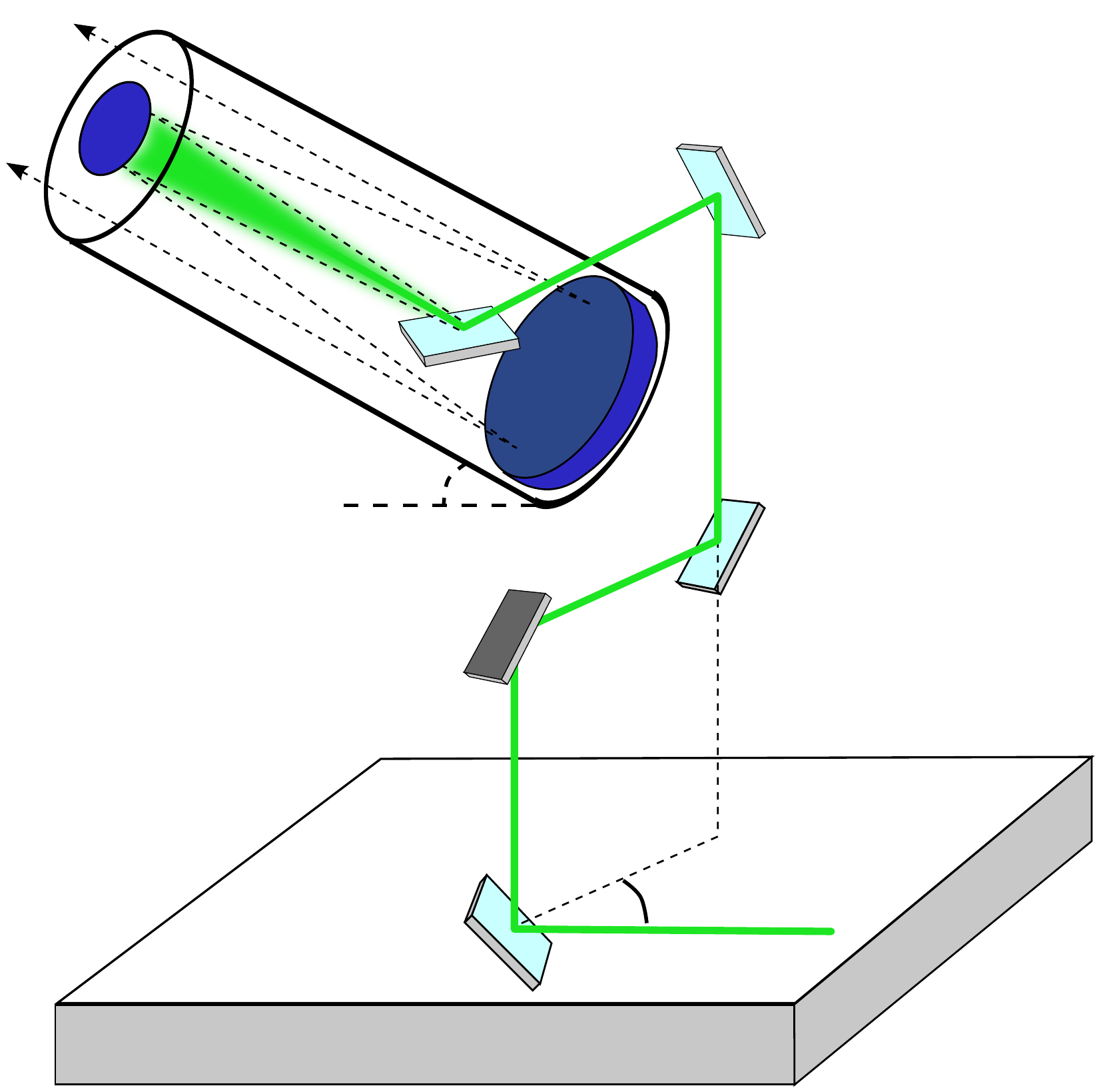}}%
	\put(0.35,0.12){\color[rgb]{0,0,0}\makebox(0,0)[lb]{$M_7$}}%
	\put(0.35,0.38){\color[rgb]{0,0,0}\makebox(0,0)[lb]{$M_6$}}%
	\put(0.68,0.46){\color[rgb]{0,0,0}\makebox(0,0)[lb]{$M_5$}}%
	\put(0.68,0.82){\color[rgb]{0,0,0}\makebox(0,0)[lb]{$M_4$}}%
	\put(0.4,0.72){\color[rgb]{0,0,0}\makebox(0,0)[lb]{$M_3$}}%
	\put(0.0,0.9){\color[rgb]{0,0,0}\makebox(0,0)[lb]{$M_2$}}%
	\put(0.48,0.62){\color[rgb]{0,0,0}\makebox(0,0)[lb]{${\color{white}M_1}$}}%
	\put(0.6,0.16){\color[rgb]{0,0,0}\makebox(0,0)[lb]{$\theta_{\rm az}$}}%
	\put(0.35,0.54){\color[rgb]{0,0,0}\makebox(0,0)[lb]{$\theta_{\rm el}$}}%
	\put(0.25,0.02){\color[rgb]{0,0,0}\makebox(0,0)[lb]{Optical table}}%
 \end{picture}%
\caption{Coud\'e path of the MLRO telescope}
\label{fig:coude}
\end{center}
\end{figure}
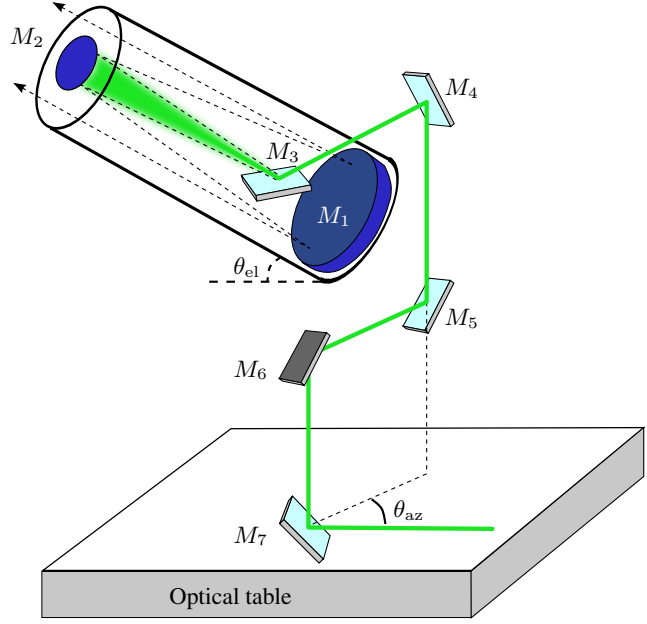

The polarization state generated on the optical table of the MLRO observatory in subjected to a unitary transformation due to the Coud\'e path of the telescope. Indeed, the Coud\'e path is composed of mirrors $M_1\,\cdots,M_7$ as in figure \ref{fig:coude}, with $M_1$ and $M_2$ the primary and secondary mirror of the telescope.
If the mirrors are coated to have $\pi$ phase shift between s- and p- polarization (corresponding to a $\sigma_z$ transformation), the transformation in the uplink channel is given by
$$U_{\rm up} = \sigma_z \, R\left(\frac\pi2-\theta_{\rm el}\right) \, \sigma _z \, R(\theta_{\rm az}) \, \sigma _z \,R\left(\frac\pi2\right)\,,$$
where $\theta_{\rm az}$ and $\theta_{\rm el}$ are the azimuth and elevation angles of the telescope and $R(\theta)$ is a rotation of the reference frame given by:
$$R(\theta)=e^{-i\theta\sigma_y}=\begin{pmatrix}\cos\theta & \sin\theta\\ -\sin\theta & \cos\theta\end{pmatrix}\,.$$
With an input polarization $\ket{\psi}=\begin{pmatrix}\cos\alpha \\ e^{i\phi}\sin\alpha\end{pmatrix}$, the polarization at the output of the telescope is given by $\ket{\psi'}=U_{\rm up}\ket{\psi}$. Since the CCR induce a transformation of $\sigma_z$ and the downlink channel can be written as
$$U_{\rm down} \,= \, R\left(\frac\pi2\right) \, \sigma _z \, R(\theta_{\rm az}) \, \sigma _z \,R\left(\frac\pi2-\theta_{\rm el}\right) \, \sigma_z\,,$$
the receiving polarization state is given by
$$\ket{\psi_{\rm rec}} \, =  \,U_{\rm down} \, \sigma_z \, U_{\rm up}\ket{\psi}\,.$$
By using the property $\sigma_z \, R(\theta) = R(-\theta) \, \sigma_z$, it is easy to demonstrate that
$$\ket{\psi_{\rm rec}} = \sigma_z \, \ket{\psi}\,,$$
showing that the uplink rotation is compensated by the downlink transformation.

This compensation is at the base of our proposed two-way protocol. Indeed, if the CCR is equipped with an active element like a Faraday Rotator at the entrance face, the transformation induced by the CCR is given by $U_{\rm CCR}(\phi) = R (-\phi) \sigma_z R(\phi) $ with $R(\phi)=e^{-i \phi \sigma_y}$. The overall transformation is then obtained as
$$\ket{\psi_{\rm rec}(\phi)}=U_{\rm down}\,U_{\rm CCR}(\phi)\,U_{\rm up}\ket{\psi}=R(2\phi)\sigma_z\ket{\psi}\,.$$
If the input state is horizontally polarized, the received state is thus rotated by an angle of $2\phi$ in the laboratory reference frame.
By modulating $\phi$, the two-way QKD protocol can be realized.

\end{document}